\documentclass[notitlepage,secnumarabic,amssymb, amsmath, tightenlines,nobibnotes, pre,longbibliography]{revtex4-1}

\usepackage[pdftex,colorlinks=true,pdfstartview=Fit]{hyperref}		

\let\oldmarginpar\marginpar
\renewcommand\marginpar[1]{\-\oldmarginpar[\raggedleft\footnotesize #1]%
	{\raggedright\footnotesize #1}}

\usepackage{color}			

\usepackage[normalem]{ulem}		

\usepackage{bm}		
\usepackage{amssymb,amsmath}
\usepackage{graphicx}

\usepackage{epstopdf}

\begin{document}
	
\title{Feedback traps for virtual potentials}
	
	\author{Mom\v cilo Gavrilov$^{1}$}
	\email[email: ]{momcilog@sfu.ca}
	
	\author{John Bechhoefer$^{1, 2}$}
	\email[email: ]{johnb@sfu.ca}
	
\affiliation{$^{1}$Department of Physics, Simon Fraser University, Burnaby, British Columbia, V5A 1S6, Canada\\
$^{2}$Kavli Institute for Theoretical Physics China CAS, Beijing, 100190, China}




\begin{abstract}
Feedback traps are tools for trapping and manipulating single charged objects, such as molecules in solution.  An alternative to optical tweezers and other single-molecule techniques, they use feedback to counteract the Brownian motion of a molecule of interest. The trap first acquires information about a molecule's position and then applies an electric feedback force to move the molecule.  Since electric forces are stronger than optical forces at small scales, feedback traps are the best way to trap single molecules without ``touching'' them.

Feedback traps can do more than trap molecules:  They can also subject a target object to forces that are calculated to be the gradient of a desired potential function $U(x)$.  If the feedback loop is fast enough, it creates a \textit{virtual potential} whose dynamics will be very close to those of a particle in an actual potential $U(x)$.  But because the dynamics are entirely a result of the feedback loop---absent the feedback, there is only an object diffusing in a fluid---we are free to specify and then manipulate in time an arbitrary potential $U(x,t)$.

Here, we review recent applications of feedback traps to studies on the fundamental connections between information and thermodynamics, a topic where feedback plays an even more-fundamental role.  We discuss how recursive maximum likelihood techniques allow continuous calibration, to compensate for drifts in experiments that last for days.  We consider ways to estimate work and heat, using them to measure fluctuating energies to a precision of $\pm 0.03~kT$ over these long experiments.  Finally, we compare work and heat measurements of the costs of information erasure, the \textit{Landauer limit} of $kT \ln 2$ per bit of information erased.  We argue that when you want to know the average heat transferred to a bath in a long protocol, you should measure instead the average work and then infer the heat using the first law of thermodynamics.
\end{abstract}
\maketitle
\section{Introduction}
In 2005, Cohen and Moerner developed a new tool for trapping and manipulating small objects in solution.  The technique traps small objects by counteracting the Brownian motion directly via electrokinetic forces that typically include electrophoretic forces (proportional to the charge of the desired object) and electroosmotic forces (proportional to the charge of ions in solution that drag the desired object).  A key advantage is that the object under study can diffuse freely in solution.  Competitive techniques all perturb the molecule more.  For example, in optical tweezers, the molecule of interest is often tethered to a long piece of DNA \cite{wang97,bustamante00,bockelmann02}.  In atomic force microscopy, the object of interest is chemically or physically bound to a substrate \cite{giessibl03,ritort06}.  In convex lens-induced confinement (CLIC), the object is confined vertically in a thin chamber \cite{leslie10}.

Feedback traps have been applied to the study of single molecules \cite{cohen06a,cohen05d,wang14,cohen07a,Goldsmith2010,Fields2011,germann14,kayci14} and to explore fundamental questions in the non-equilibrium statistical mechanics of small systems \cite{cohen05b,cohen05d,jun12,gavrilov13,Jun14,lee15,gavrilov15, gavrilov16a, gavrilov16b, proesmans16}.  For the latter applications, we have developed an improved ABEL trap that can create \textit{virtual potentials} for the particle of interest.  Reflecting this more general application and the fact that the trapping forces need not be electrokinetic---any available force will do---we adopt the more generic name of \textit{feedback trap}.

The feedback trap periodically measures the position of an object and then calculates and applies a force to keep the particle in the field of view.  The action of the feedback loop creates a virtual potential that can confine a particle or force it to perform more complicated motion.  

This use of  feedback to change qualitatively the dynamics of a system (from free diffusion to potential motion) is very much in keeping with the spirit of the ``creative interaction of physics and control theory" that has been a focus of \textit{cybernetical physics}, or \textit{cyberphysics} \cite{fradkov07}.  The tools resulting from this creative interaction have unique capabilities and have led to investigations of fundamental issues at the intersection of thermodynamics and information theory.

\subsection{Operation of a feedback trap}
\label{sec:feedbackTrapOp}

Our camera-based feedback trap is implemented around a home-built inverted microscope \cite{gavrilov13, gavrilov14, gavrilov15}.  We trap silica beads of nominal diameter 1.5~$\mu$m immersed in water.  The beads are heavy enough to sink to the bottom of the sample cell and stay in focus in the microscope but light enough to diffuse laterally.  Thus, gravity balances electrostatic repulsion from the bottom and confines particles vertically.  Two sets of electrodes control the electric field in the $x$ and $y$ directions, applying the desired force to the particle.

\begin{figure}[!h]
	\centering\includegraphics[width=4.5in]{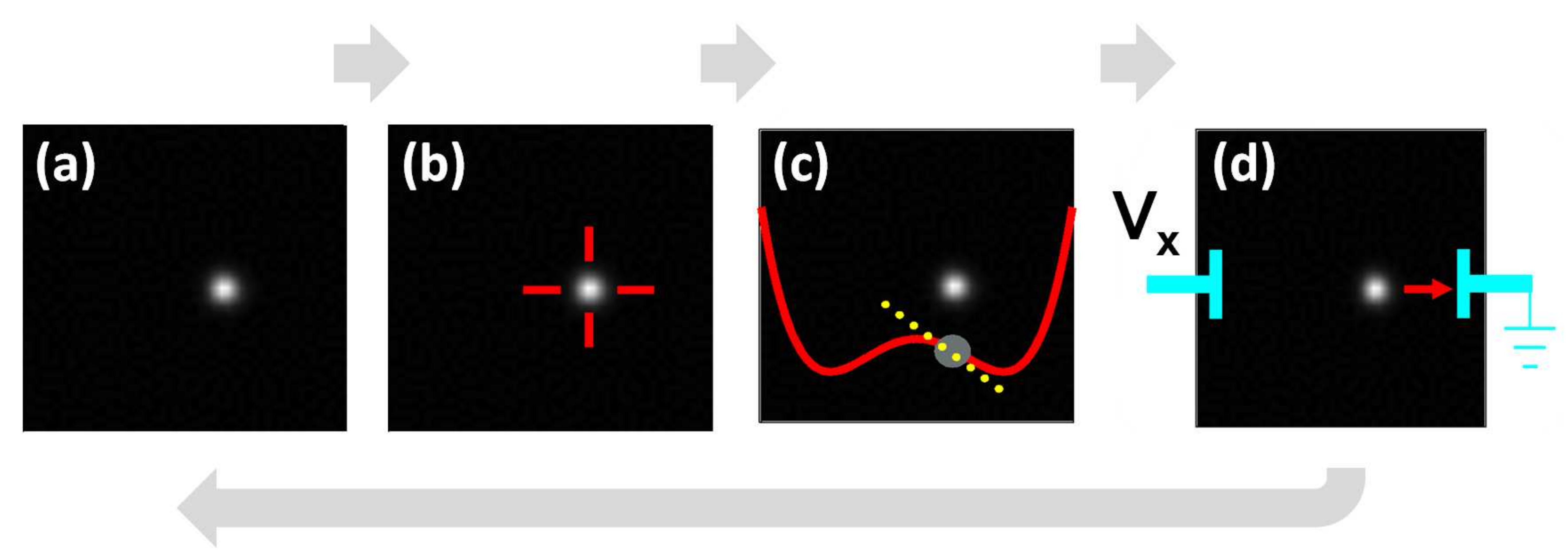}
	\caption{\label{fig:FBtrap} Schematic of feedback-trap operation.  (a) Acquisition of an image of a fluorescent particle.  (b) Determination of particle position from that image using a centroid algorithm. (c) Evaluation of feedback force   $f_x=-\partial_x U(x,t)$ at the observed position $\bar{x}$. (d) Application of electric force, with voltage set by electrodes (light blue), held constant during the update time $\Delta t= 5$ ms. The long gray arrow indicates repetition of the cycle. Figure reproduced from ref.~\cite{Jun14}.}
\end{figure}

Figure~\ref{fig:FBtrap} shows one cycle of a feedback trap in operation  \cite{gavrilov13,gavrilov14}.  The camera first images a particle using a dark-field microscope.  From the image, the particle position $\bar{x}_n$ is estimated using a centroid algorithm \cite{berglund08}, and the feedback force is calculated from the chosen potential \mbox{$f_x=-\partial_x U(x,t)$}, evaluated at the observed position $\bar{x}_n$.  Finally, the feedback trap applies a voltage $V_n$ that is calibrated to correspond to the desired force $f_n$, which moves the particle.  The voltage is kept constant for $\Delta t= 5$~ms and then updated with a delay $t_d= 5$~ms relative to the exposure midpoint.

In the original version of the ABEL trap \cite{cohen05b} and in our first version \cite{gavrilov13,gavrilov14}, as well, the imaging was based on fluorescence illumination, as is appropriate for fluorescent particles and for labeling individual molecules in solution.  In such low-light conditions, camera-based imaging systems need to integrate all the time to capture as much information from a fluorescent particle as possible, leading to a camera-exposure time $t_c = \Delta t$.  Even with such a long exposure, the overall light levels are low, implying significant observation noise due to photon shot noise and the diffraction limit of the microscope.  In the much brighter scattering-based illumination described above, the camera time can be short ($t_c/\Delta t \le 0.1$) and the measurement noise can be small \cite{gavrilov15}.  The simplified theoretical description developed below takes advantage of this improved illumination scheme.

\section{Stochastic dynamics in a feedback trap}

In this paper, we focus on applications of the feedback trap to the study of small particles in virtual potentials whose form we can control and manipulate at will.  Because the objects are small, their fluctuations are important, and describing energy terms requires the formalism of \textit{stochastic thermodynamics} \cite{sekimoto10,seifert12,VdBroeck13}, which considers fluctuating mesoscopic systems.  In this case, the system (the particle) is immersed in a heat bath of constant temperature, so that all processes are isothermal.  Below, we outline some of the relevant aspects that show how the trap dynamics are related to those of the feedback loop and how both are related to thermodynamic quantities.  

%

\subsection{Trap dynamics}
The low-frequency stochastic dynamics of a Brownian particle are well described by a Langevin equation in the overdamped limit.  For a one-dimensional trajectory $x(t)$, the Langevin equation is
\begin{align}
	\gamma \, \dot{x} &= f(x, t) + \xi^{(f)}(t) \, ,
\label{eq:LangevinEqIntro}
\end{align}
where $\gamma$ is the drag coefficient, $\xi^{(f)}(t)$ is the fluctuating thermal force acting on the particle, and $f(x, t)$ is the external force.  Here, the force $f(x, t) = -\partial_x U(x,t)$ is chosen to be the gradient of a time-dependent potential $U(x, t)$.  In equation~\ref{eq:LangevinEqIntro}, we have neglected the mass term, $m\ddot{x}$, as the inertial time scale, $m/\gamma$, is of order microseconds, whereas experimental time scales are of order milliseconds.


Although equation~\ref{eq:LangevinEqIntro} is one dimensional, the particle moves in three dimensions.  We have already discussed how gravity and electrostatic forces naturally confine the motion in $z$.  For the lateral $y$-direction, we confine the particle by adding an additional harmonic component, so that $U(x,y,t) = U(x,t) + \tfrac{1}{2}\kappa y^2$.  The $y$ motion plays no role in work and heat estimates and is ignored in the rest of the analysis, which focuses on the one-dimensional motion along $x$.

To discretize equation \ref{eq:LangevinEqIntro}, we integrate it over the time interval $\left [t_n, t_{n+1} \right)$, where $t_n\equiv n\Delta t$ and $\Delta t$ is the feedback cycle time.  The force is kept constant in that time interval:  \mbox{$f(x,t)\equiv f_{n-1}$}, for $t \in \left[ t_n,t_{n+1} \right)$.  The discretized position of the particle ($x_n$) at the beginning of that interval ($t_n$) is
\begin{equation}
	x_{n+1} = x_n + \tfrac{\Delta t}{\gamma}f_{n-1} + \xi_n\, ,\qquad  \qquad
		\xi_n = \frac{1}{\gamma} \int_{t_n}^{t_{n+1}} dt \, \xi^{(f)}(t) \,.
\label{eq:ActualPosition}
\end{equation}
In equation~\ref{eq:ActualPosition}, $\xi_n$ represents the displacement due to thermal forces that are integrated over the cycle time $\Delta t$ of a single time step.  The displacements are described by i.i.d. random Gaussian variables with mean 0, and variances $\left < \xi_n^2 \right > = 2D \, \Delta t $, with $D$ the lateral diffusion coefficient of the particle.  As discussed above, in general, we would supplement equation~\ref{eq:ActualPosition} with a second equation for the observed position, which incorporates effects due to exposure averaging, measurement noise, and the rather subtle noise correlations that follow \cite{jun12}.  Fortunately, these are all small effects in the current setup and will be ignored here.

To simplify the equation of motion, let us introduce coordinates where lengths are scaled by $\ell = \sqrt{D\Delta t}$ and energies by $kT$.  We also use the Einstein relation $D = kT/\gamma$.  More formally, we define (omitting the time subscript for simplicity),
\begin{align}
	x' = \frac{x}{\ell} \,, \qquad \xi' =  \frac{\xi}{\ell} \,, \qquad U' = \frac{U}{kT} \,, \qquad
		f' = \frac{\ell f}{kT} \,, \qquad t' = \frac{t}{\Delta t} \,.
\label{eq:scalings}
\end{align}
In terms of these scalings, equation~\ref{eq:ActualPosition} becomes
\begin{equation}
	x'_{n+1} = x'_n + f'_{n-1} + \sqrt{2} \, \xi'_n \,, \qquad \quad
	f'_{n-1} = - \left. \frac{\partial U'(x',t')}{\partial x'} \right|_{x' \, = \, x'_{n-1}, \, t' \, = \, n-1} \,.
\label{eq:ActualPositionScaled}
\end{equation}
Note the $n-1$ subscript for the force term, which reflects the unit delay of the feedback system.  In our experimental design, we take care to make sure that the updated force is applied precisely at a time $\Delta t$ after the midpoint of the exposure, as this is necessary in the simplified equations we give here.  We note that the experiment is particularly sensitive to timing variations and that using ``real time'' programming techniques and hardware is important.  Finally, in equation~\ref{eq:ActualPositionScaled}, the scaled noise terms $\xi'_n$ are independent Gaussian random variables with $\langle \xi'_n \rangle = 0$ and $\langle \xi'_n{}^2 \rangle = 1$.

An important special case is a harmonic potential, $U = \tfrac{1}{2} \kappa x^2$, or $U' = \tfrac{1}{2} \alpha x'^2$, where \mbox{$\alpha = \kappa \ell^2/ kT = \kappa \Delta t / \gamma = \Delta t/ t_r$}.  In the last version, $t_r$ is the relaxation time of an overdamped particle in a harmonic potential.  More generally, we can replace $\kappa$ by $\partial_{xx} U(x,t)$, evaluated at the bottom of a local potential well.  For a harmonic potential, the scaled equation of motion is
\begin{equation}
	x'_{n+1} = x'_n -\alpha \, x'_{n-1} + \sqrt{2} \, \xi'_n \,.
\label{eq:harmonic}
\end{equation}
In earlier work, we explored in detail the dynamics and thermodynamics of a particle in a virtual harmonic potential, taking into account the delay, exposure time, and the observational noise \cite{jun12}.  If the relaxation time of the potential is much greater than the update time of the feedback trap $t_r\gg\Delta t$, or $\alpha = \Delta t/ t_r \ll 1$, the dynamics approaches the dynamics of a particle in a physical potential.  In our experiment, the feedback gain never exceeds $\alpha=0.2$.  At such values of $\alpha$, one might worry that corrections begin to be significant.   Nevertheless, for a cyclic protocol, where the potential is the same at the beginning and the end of a protocol, the correction term cancels out, and the work measurement is exact \cite{gavrilov16b}.

\subsection{Trap calibration}

The energy scale of the processes we consider are of order $kT \approx 4 \times 10^{-21}~J$, where $T$ is the temperature of the heat bath the system is immersed in and $k$ is Boltzmann's constant.    For detecting such small energy changes in protocols that can last minutes, we need careful, high-precision calibration techniques that can convert the applied voltages into precisely known forces \cite{gavrilov14}.  Here, we briefly summarize the calibration process.

Rearranging equation \ref{eq:ActualPositionScaled} for the feedback trap dynamics, we define the  displacement \mbox{$\Delta x'_n \equiv x'_{n+1} -x'_n = f'_{n-1} + \sqrt{2} \, \xi'_n$}.  Thus, the displacement results from a deterministic component, $f'$, which we control, and a stochastic component, $\xi'$, which we do not.  To control the forces $f'_{n-1}$, we apply a voltage $V_{n-1}$, which generates an electric field that moves the charged particle to a new position.  In other words, the controllable component of displacement $f'_{n-1}$ is proportional to the applied voltage as \mbox{$f'_{n-1} = \mu \, \Delta t \, (V_{n-1} + V_{\rm{o}})$}, where $\mu$ is the mobility (response) of a charged particle (in an ionic fluid) to the applied voltage.  The offset $V_{\rm{o}}$ arises from electrochemical reactions at the electrodes and from the amplifier in the circuit.  The link between displacement $\Delta x'_n$ and applied voltage is then 
\begin{equation}
	\Delta x'_n= \mu \, \Delta t \, V_{n-1}+\mu \, \Delta t V_{\rm{o}} + \sqrt{2} \, \xi'_n \, .
\label{eq:LangevinVoltage}
\end{equation}

In principle, we could estimate $\mu$ and $V_{\rm{o}}$ as the slope and intercept in a linear, least-squares fit to $N$ measurements and could similarly infer the diffusion coefficient $D$ from the residuals.  However, our goal is to implement a continuous calibration, and we cannot afford to recompute such a fit at every time step in the experiment.  Fortunately, it has long been known that linear least-squares fits have a \textit{recursive} formulation where existing estimates of slope and intercept are updated each time a new data point arrives \cite{Astrom2008}.  Here, in order to treat small noise correlations (arising from term neglected in equation~\ref{eq:ActualPosition}), we use a slight generalization known as \textit{Recursive Maximum Likelihood} (RML) \cite{Astrom2008,gavrilov14}, which is a variant of the \textit{Kalman filter} \cite{bechhoefer2005}.  

One complication is that experimental parameters slowly drift due to temperature drifts and electrochemical reactions.  We thus limit the amount of the past time series measurements that enter into the calibration by using a \textit{running average} variant of the RML algorithm that weights new versus old measurements \cite{gavrilov14}.

Another complication is that the feedback trap corrects motion in two dimensions.  Since the electric-field lines are not exactly aligned along the $x$-$y$ axes (defined by the camera pixels), we need to implement a full two-dimensional version of the RML algorithm that accounts for coupling between axes.  Apart from generalizing the mobility $\mu$ to a two-by-two matrix $\bm{\mu}$ and the offset voltage to a vector $\mathbf{V}_0$, the algorithm is identical.


\section{From dynamics to thermodynamics}

In an important conceptual advance for extending thermodynamics to small systems, Sekimoto proposed a method for determining the stochastic energetics of a particle solely from its Langevin dynamics \cite{sekimoto97,sekimoto10}.  He showed that thermodynamical quantities such as  work and heat can be estimated solely from a particle's trajectory $x(t)$ and the shape of the potential $U(x, t)$.  By using just the trajectory and the potential, one can isolate and measure the quantities of direct interest.  This method removes the contributions of work and dissipation from ancillary devices---computer, camera, illumination, etc.---that are irrelevant to calculating the work done by the potential on the particle and the heat dissipated into the surrounding bath.

\subsection{Work and heat and the first law}

The first law of thermodynamics can be formulated for an individual fluctuating trajectory:
\begin{equation}
	dW = dE + dQ \,,
\label{eq:FirstLaw}
\end{equation}
where $dW$ is the work increment done by the changing potential on the particle, $dQ$ is the heat transferred to the environment by the motion of the particle, and $dE$ is the change in the potential energy of the particle in the external potential.  Equation \ref{eq:FirstLaw} uses a sign convention where \mbox{$dW>0$} refers to work done by the changing potential on the particle and \mbox{$dQ>0$} means that heat is transferred to the bath.  This convention is convenient for comparing work and heat in a cyclic process, as both distributions are centered on the same mean.

As proposed by Sekimoto \cite{sekimoto97,sekimoto10}, the work over the entire trajectory is simply the sum over potential changes evaluated at the local position of the particle, and the heat is the sum of force times velocity:
\begin{equation}
	W = \int_0^\tau dt\, \left. \frac{\partial U(x, t)}{\partial t} \right|_{x=x(t)} \,, \qquad \quad
	Q = \int_0^\tau dt  \, \dot{x} \, \left. \frac{\partial U(x, t)}{\partial x} \right|_{x=x(t)} 
		= - \int_0^\tau dt  \, \dot{x} \, f[x(t),t] \,.
\label{eq:WorkHeat}
\end{equation}
Because the trajectory is fluctuating and different for each realization of the experimental protocol, so are the values of $W$ and $Q$; nevertheless, the first law applies both to the individual trajectory and to the ensemble average over those trajectories:
\begin{equation}
	\underbrace{W  = Q + \Delta E}_{\rm stochastic} \quad \Longrightarrow \quad
	\underbrace{\left < W \right > = \left <Q \right > +\left < \Delta E \right>}_{\rm ensemble} \,.
\label{eq:FirstLawStochEns}
\end{equation}

\subsection{Numerical estimates of work and heat}

Equation \ref{eq:WorkHeat} gives a way to estimate work and heat from a single continuous trajectory of a particle in a physical potential $U(x,t)$.  There are two issues in applying it to our experiments.  First, our trajectories are discrete and not continuous.  Second, as equation~\ref{eq:ActualPositionScaled} indicates, the dynamics of a particle in a virtual potential generally differ from those of a particle in the corresponding physical potential.  For example, changing the stiffness of a harmonic potential quasistatically leads to expected work values of $\langle W \rangle~=~\Delta F~+~ \mathcal{O}(\alpha)$, where $\alpha \equiv \Delta t / t_r$ is defined above equation~\ref{eq:harmonic} and where $\Delta F$ is the change in equilibrium free energy \cite{jun12}.  We note, however, that cyclic operations such as expanding and then compressing lead to $\langle W \rangle = 0$, exactly \cite{gavrilov16b}.

The naive discretization of the stochastic work is simply
\begin{equation}
	W (\tau) \approx \sum_{n=0}^N \left.
		\frac{\partial U'(x'_n, n)}{\partial t'} \right|_{x' \, = \, x'_n, \, t' \, = \, n} \,, \label{eq:WorkDiscrete}
\end{equation}
where we have omitted the term $\Delta t' = 1$ from the sum and where $N = \tau / \Delta t$.

The discretization of heat is more subtle.  Because $\dot{x}$ appears in its definition, we can use the Langevin equation to write 
\begin{equation}
	Q(\tau) \approx -\frac{1}{\gamma} \int_0^\tau dt  \, 
		\left\{ f[x(t), t] + \xi^{(f)}(t) \right\} \, f[x(t),t] \,.
\label{eq:Heat1}
\end{equation}
In equation~\ref{eq:Heat1}, we see that there is now multiplicative (state-dependent) noise:  the amplitude of $\xi^f(t)$ is multiplied by $\partial_x U = f(x,t)$.  Discretization then leads to the well-known ``Ito-Stratonovich'' dilemma, which stems from the need to choose a point in the time interval $\Delta t$ to evaluate the force $f$ \cite{vanKampen92}.  In standard discussions of stochastic dynamics, it is assumed that the Stratonovich convention is the appropriate one, where the forces that control the noise amplitude are evaluated at the midpoint of the time interval.  This leads to
\begin{equation}
	Q(\tau) \approx - \sum_{n=1}^N \left( x'_n - x'_{n-1} \right) \, 
		 f'\left( \frac{x'_n+x'_{n-1}}{2},n-\tfrac{1}{2} \right)  \,.
\label{eq:HeatDiscrete}
\end{equation}

Unfortunately, the discretization proposed in equation~\ref{eq:HeatDiscrete} is problematic.  To see this, consider a test case where it is satisfactory, the static harmonic potential.  In dimensionless units, the force $f' = -\alpha x'$.  Because the potential is not altered, the stochastic work $W$ is identically zero.  Then

\begin{equation}
	Q(\tau) \approx -\sum_{n=1}^N \left( x'_n - x'_{n-1} \right) \, 
		\alpha \tfrac{1}{2} \left( x'_n + x'_{n-1} \right) 
	= -\tfrac{1}{2}\alpha\sum_{n=1}^N \left( {x'_n}^2 - {x'_{n-1}}^2 \right) = \tfrac{1}{2}\alpha{x'_0}^2 - \tfrac{1}{2}\alpha{x'_N}^2 \,.
\label{eq:HeatDiscreteQuadratic}
\end{equation}
The last term in equation~\ref{eq:HeatDiscreteQuadratic} is just $U'_0 - U'_N = -\Delta E$.  Thus, both \mbox{$Q = -\Delta E$} and \mbox{$\left< Q \right> = -\left< \Delta E \right>$}, as required by equation~\ref{eq:FirstLawStochEns}.

Next, we try $U(x') = \tfrac{1}{4} {x'}^4$, or $f' = {x'}^3$.  Evaluating $f'$ at the midpoint then gives 
\begin{align}
	Q(\tau) &\approx -\sum_{n=1}^N \left( x'_n - x'_{n-1} \right) \, 
		\tfrac{1}{2^3} \left( {x'_n} + {x'_{n-1}} \right)^3 \nonumber \\
	&= -\frac{1}{8} \sum_{n=1}^N \left( {x'}_n^2 - {x'_{n-1}}^2 \right) \left( {x'_n} + {x'_{n-1}} \right)^2
		\nonumber \\
	&= -\frac{1}{8} \sum_{n=1}^N 
	\underbrace{{x'}_{n+1}^4-{x'}_n^4}_\text{cancels}
	+ \underbrace{2 x'_n x'_{n+1} \, ({x'}_{n+1}^2-{x'}_n^2)}_\text{does not cancel} \,.
\label{eq:HeatDiscreteQuartic}
\end{align}
The terms that cancel lead to $\Delta E$, as before; however, the terms that do not cancel give an error at each time step.  In a long time series, this \textit{secular} error dominates, making the estimator useless.

The heart of the difficulty, that the numerical method does not respect conservation of energy, is a familiar one in numerical analysis, and methods such as \textit{velocity Verlet} and \textit{symplectic integration} have been developed for solving this problem in Hamiltonian systems \cite{blanes16}.  For Langevin equations, Sivak et al. \cite{Sivak13} derived a similar scheme and showed that it made such secular errors very small.  However, those methods were derived for systems with momentum variables and break down in the Brownian-motion limit, where mass terms are taken to zero.

In private correspondence, D.~Chiuchi{\`u} noted that the naive estimator is equivalent to a trapezoidal integration scheme and suggested substituting a higher-order integrator such as Simpson's Rule.  As he emphasized, from general discussion of the integration of stochastic differential equations, such a scheme cannot be quite correct because one must consistently handle both the deterministic and stochastic terms.  Yet we will see below that the scheme works in the sense that numerical estimation errors do not increase linearly in time.

Adapting Simpson's Rule\footnote{
For an interval $\Delta x$, Simpson's rule is \mbox{$\int_x^{x+\Delta x} dx'\, f(x') \approx \Delta x \, \left[ \tfrac{1}{6} f(x) + \tfrac{4}{6} f(x+\frac{1}{2}\Delta x) + \tfrac{1}{6} f(x+\Delta x)\right ]$.}} 
to the estimation of the heat gives
\begin{align}
	Q(\tau) \approx  -\sum_{n=1}^N 
	\left( x'_{n+1} - x'_{n} \right) \, \left [ \tfrac{1}{6} f' \left( x'_n,n \right) 
	+\tfrac{4}{6} f' \left( \frac{x'_n+x'_{n-1}}{2},n-\tfrac{1}{2} \right) + \tfrac{1}{6} f' \left( x'_{n-1},n-1 \right)  \right] \,.
\label{eq:simpsonHeat}
\end{align}

\section{Experimental test of the Landauer principle}
The Landauer principle, formulated in 1961, connects the seemingly disparate concepts of information, work, and heat exchange.  Landauer argued that \textit{information is physical}---it is tied to a physical representation and therefore restricted to what the laws of physics allow \cite{Landauer61, landauer96}.  

According to this physical view of information and its processing, erasing information in a macroscopic or mesoscopic system should require a minimum amount of work, $kT \ln 2$ per bit erased.  The heat in this process is transferred to the bath, reversibly if the erasure is done slowly enough and with a properly chosen protocol \cite{gavrilov16a}.  At the time, the immediate motivation was to understand the minimum power a computer requires to function.  Later, Bennett recognized the relevance of Landauer's principle in explaining paradoxes raised by the Maxwell-demon thought experiment \cite{bennett73,bennett82}.  In particular, Szilard showed, in a slight adaptation of Maxwell's original scenario \cite{szilard29}, that acquiring one bit of information about a system allows the extraction of $kT \ln 2$ of work from the surrounding heat bath, in apparent violation of the second law of thermodynamics.  Landauer's principle then implies that the same amount of work (or more) must be done when erasing the acquired information. 

Modern theoretical developments have expanded this picture into one where \textit{measurement} correlates a physical measuring device with a physical system of interest.  \textit{Feedback} then uses these correlations to extract work from the physical system.  Finally, the measuring device is \textit{reset} to complete a full cycle where system and measuring device are in their original states.   Thus, in a complete cycle, the net amount of work done is zero, or more.  The second law thus holds for the combined physical and measurement systems, also known as \textit{information engines} \cite{parrondo15}.  
  
A number of recent experiments have tested various aspects of this picture \cite{toyabe10, roldan14, martinez15,blickle12,koski13, koski14, koski14b, koski15}.  Here, we focus on studies of the erasure (reset) process, which have tested the Landauer principle by measuring either the heat released into the surrounding bath \cite{berut12} or the work done \cite{Jun14} to erase a one-bit memory.  In the results reported in this article, we simultaneously measure both the work required and the heat transferred to the bath in protocols that erase a one-bit memory.  Simultaneous work and heat estimates have been compared and analyzed in theoretical studies\cite{dillenschneider09, chiuchiu15}, but not in experiments \cite{Hong16, Martini16, Peterson16, rossnage16, berut12, Jun14, berut13}.  Here, we compare distributions of these stochastic quantities for an erasure experiment and consider the relative merits and limitations of each type of measurement.

\subsection{Virtual double-well potential}

We have adapted the feedback trap described above to study memory erasure in a double-well potential.  In one dimension, the potential is parametrized to control the height of the barrier $E_b$ and tilt $A$ independently \cite{dillenschneider09,chiuchiu15,jun12,gavrilov16b}:
\begin{equation}
	U(\tilde{x},t) = 4 E_b \left[ -\tfrac{1}{2} g(t) \, \tilde{x}^2 
		+ \tfrac{1}{4} \tilde{x}^4 - A h(t) \, \tilde{x} \right] \,,
\label{eq:DWpotential}
\end{equation}
where $\tilde{x} \equiv x/x_m$ is a (re)scaled coordinate with $x_m=0.7$~$\mu$m.  The barrier height at the beginning and end of each erasure cycle is set to $E_b/kT=12$, and the maximum tilt amplitude $A=0.2$.  The functions $\{ g(t), \, h(t) \} \in [0,1]$ alter the barrier height and tilt, respectively.  They are chosen to define a thermodynamically reversible protocol, while being more efficient than the protocol that we used in \cite{jun12}.  The efficiency gain comes from starting the tilt before the barrier has fully descended.  One must be careful not to start tilting too early, as irreversibility can occur if probability fluxes are not matched when the probability densities in the two regions of the well begin to mix \cite{gavrilov16a,gavrilov16b}.  An important property of the potential in equation \ref{eq:DWpotential} is its curvature around the well minima, which can be evaluated as
\begin{equation}
\left| \kappa_m \right| = \left| \frac{d^2 \, U(x)}{d  x^2} \right|_{x=\pm x_m} 
	= 8 \frac{E_b}{(x_m)^2}\,.
	\label{eq:curvature}
\end{equation}
The relaxation time of a particle within the basin of one well is $t_r=kT/(\kappa_m D)$.

\begin{figure}[!h]
	\centering\includegraphics[width=4.5in]{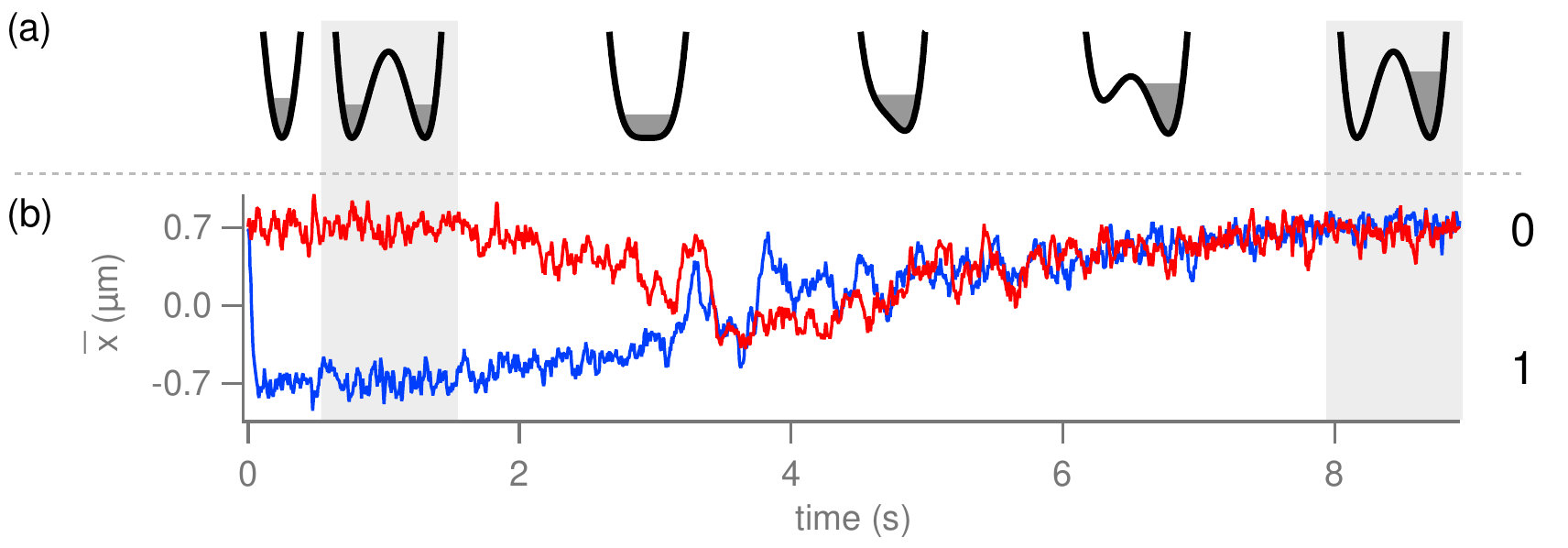}
	\caption{\label{fig:ObsPos} Observed trajectories and the virtual trapping potential in the erasure experiment.  (a) The erasure protocol with the initial state preparation.  Shaded strips indicate the static double-well potential applied during 1~s.  During this time, the applied work is zero, but heat is continuously exchanged with the bath.  (b) Two observed stochastic trajectories for two initial states.  All trajectories end up in the right well.}
\end{figure}

Figure \ref{fig:ObsPos}a illustrates the evolution of a virtual potential under the protocol used in the erasure experiment.  The erasure experiment starts from the equilibrium state.  The harmonic trap is applied for 0.5~s, and then the potential is abruptly changed to a static double well (equation \ref{eq:DWpotential}).  Since the harmonic trap position corresponds to the position of one or the other well in the new potential, the perturbations due to the potential switch are small.  The static double-well potential is applied for 1~s to allow particle to relax within the chosen well.  The relaxation time within one well is at the order of 30~ms.  The initial conditions are chosen so that there are equal numbers of 0 and 1 states.

As the sketches in Figure \ref{fig:ObsPos}a show, the erasure protocol first lowers the barrier to mix states.  After the barrier has been removed, the logical degrees of freedom are returned to the bath.  We further start tilting towards $\tilde{x}=1$, followed by raising the barrier to recreate logical states.  In the end, potential is untilted and returns to its original shape.  In order to accurately estimate heat in the erasure experiment, we let the particle relax in a static double-well potential for 1~s at the end of the erasure cycle.  This step is not necessary for the work estimate.

Figure \ref{fig:ObsPos}b illustrates two complete trajectories for a particle starting in each well.  Each trajectory shows the initial harmonic trap, relaxation in a static double-well potential (gray shaded area), and the erasure protocol.    

The experiment is repeated for several different vales of the scaled erasure cycle time from $\tau=0.33$ to $\tau=25$.  The cycle time in seconds $t_{\rm cyc}$ is linked to the scaled cycle time as $\tau \equiv t_{\rm cyc} / \tau_0$, where the scaling factor is $\tau_0~=~(2x_m)^2/D \approx9$~s.

\subsection{Work and heat during erasure}

To estimate the work and heat, we adapt equations~\ref{eq:WorkDiscrete} and \ref{eq:simpsonHeat} to the erasure protocol for the double-well potential.  For the work, defining 
$(\Delta g)_n \equiv  \dot{g}(t_n) \, \Delta t$ and $(\Delta h)_n \equiv \dot{h}(t_n) \, \Delta t$, we have 

\begin{equation}
	W (\tau) = - 4 E_b \sum_{n=0}^N  
	\left[ \tfrac{1}{2} (\Delta g)_n \, \tilde{x}_n^2  + A (\Delta h)_n \, \tilde{x}_n \right] \,,
\label{eq:WorkCalculation}
\end{equation}

Similarly, for the heat, we follow equation~\ref{eq:simpsonHeat}, with 
\begin{equation}
	f'(\tilde x, t) = \frac{4E_b}{x_m}\left[ -g(t)\tilde{x} + \tilde{x}^3  - h(t)A \right].
\end{equation}
A second consideration for heat calculations is the need to add extra time steps to the end of the protocol, to ensure that the system relaxes to equilibrium.  This differs from the work calculation, where all contributions to the work are linked to the changing potential and cease immediately at the end of the protocol.  In practice, we found that waiting 0.5 s was long enough (by adding 100 time steps), since the relaxation time within one well is $t_r\approx0.03$~s.

\begin{figure}[!h]
\centering\includegraphics[width=4.5in]{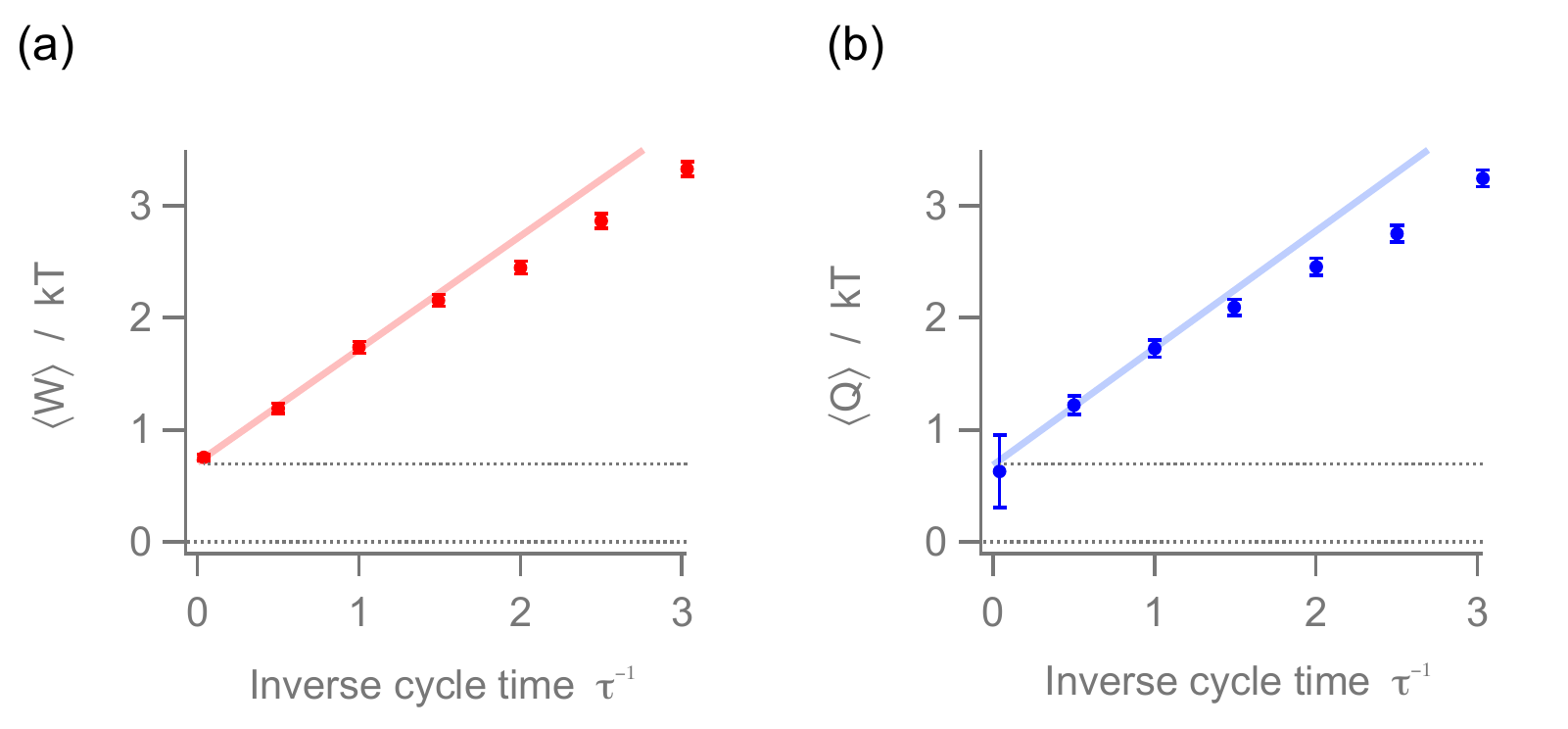}
\caption{Average work and heat to erase one-bit memory.  Solid lines shows fit to the asymptotic correction, $\tau^{-1}<1$. The asymptotic work $\langle W\rangle_\infty$ is the y-intercept in (a). The asymptotic heat $\langle Q\rangle_\infty$ is the y-intercept in (b).}
\label{fig:WandQ} 
\end{figure}

Figure \ref{fig:WandQ} shows the average work and heat measured in the erasure experiment for different values of the inverse cycle time $\tau$.  

We estimate work and heat in the arbitrarily slow limit by fitting lines to the expected asymptotic forms \cite{sekimoto97a} 
\begin{align}
	\langle W \rangle_\tau/kT \sim \langle W\rangle_\infty /kT + a_w  \tau^{-1}\,, \qquad			\langle Q \rangle_\tau/kT \sim \langle Q\rangle_\infty /kT + a_q  \tau^{-1} .
\label{eq:asymWQ}
\end{align}
From the data presented in figure~\ref{fig:WandQ}, the bound on work to erase a one-bit memory is \mbox{$ \langle W\rangle_\infty /kT = 0.71 \pm 0.03$}, while the heat released to the bath is \mbox{$ \langle Q\rangle_\infty /kT = 0.69 \pm 0.16$}.  Both measurements are compatible with the expected Landauer bound of $kT \ln 2$, but the uncertainty in the heat measurement is five times greater.  We note in figure~\ref{fig:WandQ} that the fits are limited to $\tau^{-1} < 1$ (long times), in order for the asymptotic form in equation~\ref{eq:asymWQ} to be valid.  Deviations for $\tau^{-1} > 1$ (short times) are expected to be protocol dependent.

\subsection{Work and heat distributions}

\begin{figure}[!h]
	\centering\includegraphics[width=5in]{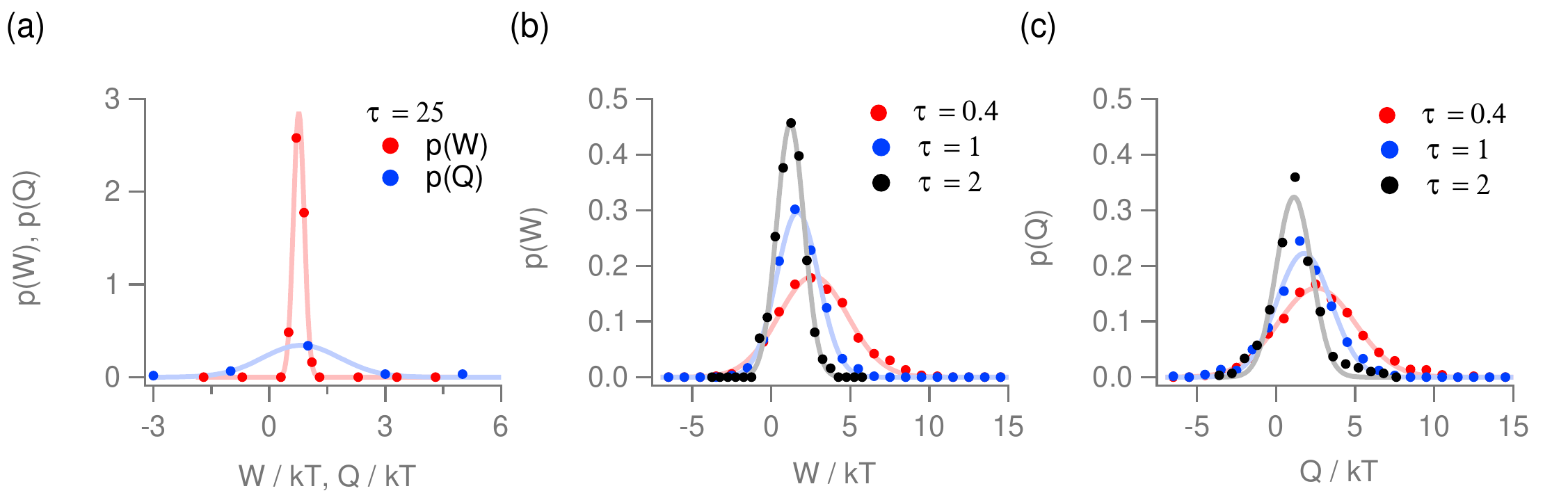}
	\caption{\label{fig:pWpQ} Work $p(W)$ and heat $p(Q)$ distributions.  (a) Comparison between work and heat distributions for a long erasure cycle time, $\tau=25$. (b) Work distributions for shorter cycle times. (c) Heat distribution for the same cycle times.}
\end{figure}

The greater uncertainty in the heat measurement might seem surprising, since both work and heat measurements are based on the same data.  To understand why the work estimate has smaller statistical errors, we first look at the individual work and heat distributions.  Figure \ref{fig:pWpQ}a shows those distributions for the cycle $\tau = 25$, where the width of $p(W)$ is narrower than the width of $p(Q)$, implying variances $\sigma_Q^2 >\sigma_W^2$ \cite{dillenschneider09}.  Figure \ref{fig:pWpQ}b and c compares work and heat distributions for different cycle times.

For the slow protocols considered in these experiments, we can derive an approximate relation between the variance of work and heat.  Starting from the first law, we subtract $\Delta E$ from equation~\ref{eq:FirstLawStochEns}, square both sides, and find the mean over the ensemble of trajectories.  Substituting \mbox{$\langle y^2\rangle = \sigma_y^2 + \left < y  \right >^2$} for $y = W, \Delta E, Q$ then gives 
\mbox{$\sigma_W^2 + \sigma_{\Delta E}^2 = \sigma_Q^2 + 2 \left[ \left < W \Delta E\right > 
- \left < W\right > \left < \Delta E\right > \right]$}.  For a cyclic erasure protocol, there is no change in potential energy.  The measurement fluctuates around a zero mean; hence, $\left <\Delta E\right >=0$.  At the end of each erasure cycle, we let the particle relax in a static double well.  As a result of this protocol feature, the energy variation $\Delta E$ loses any correlation with the work measurement $W$, making \mbox{$\left < W \Delta E\right > \approx 0$}.  Putting all these features together, we relate the variances of work, heat, and the change in potential energy:
\begin{equation}
	\sigma_W^2 + \sigma_{\Delta E}^2 \approx \sigma_Q^2 \,.
\label{eq:variances}
\end{equation}
We emphasize that equation~\ref{eq:variances} is written for our particular experiment and  protocol.  In general, there will be correlations between $W$ and $\Delta E$.

\subsubsection{Fluctuations in work and its distribution}
The Jarzynski equality describes fluctuations in small systems \cite{jarzynski97,jarzynski11}.  It generalizes the Clausius inequality ($W\geq \Delta F$) to the case of mesoscopic systems.  Unlike the Clausius inequality, the Jarzynski equality gives an exact relationship between fluctuating work $W$ and the equilibrium free energy change in a system $\Delta F$.  The Jarzynski equality was later generalized to include links between stochastic work and the \textit{nonequilibrium free energy} change $\Delta \mathcal{F}$ from an equilibrium starting state to a final state that can be far from equilibrium \cite{parrondo15}:

\begin{equation}
	\left < e^{-W/kT} \right > = e^{-\Delta \mathcal{F}/kT}  \, .
\label{eq:JarzynskiIntro}
\end{equation}

The Jarzynski equality allows one to estimate the free energy difference $\Delta \mathcal{F}$ from the non-equilibrium work measurements.  In our case, the erased memory remains out of equilibrium for a time much longer than the duration of the erasure experiment; consequently, we use the nonequilibrium version of the Jarzynski equality.  Because the exponential average in equation~\ref{eq:JarzynskiIntro} is dominated by the rare values on the left tail of the work distribution \cite{jarzynski06}, it requires large amounts of data for a proper estimate.  But for certain classes of work distributions, equation~\ref{eq:JarzynskiIntro} can be simplified.  For example, if the work distribution is Gaussian, as it is here, then \cite{seifert12}
\begin{align}
	\sigma_W^2 = 2 \left( \langle W \rangle - \Delta \mathcal{F} \right)\,.
\label{eq:GaussCumulantIntro2}
\end{align}

\begin{figure}[!h]
	\centering\includegraphics[width=5in]{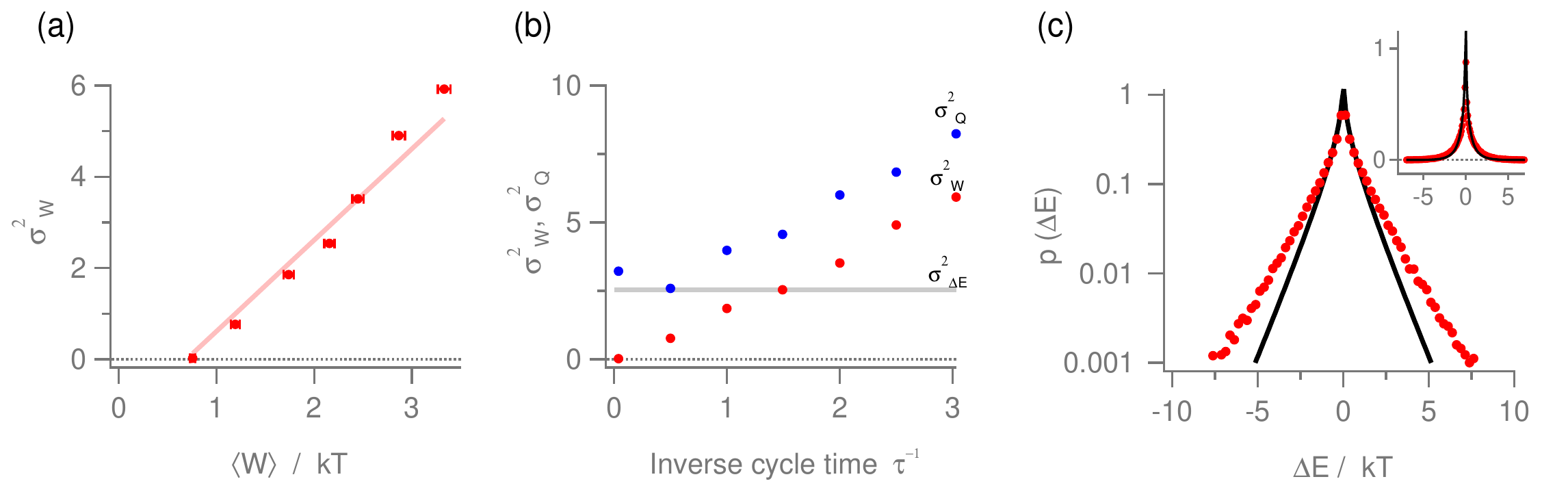}
	\caption{Fluctuations in work, heat, and energy. (a) Work fluctuations are consistent with the  Jarzynski equality, as described by equation~\ref{eq:GaussCumulantIntro2} (solid red line) (b) Variance of heat measurements exceeds those of work measurements at all cycle times.  Solid gray line shows measured $\sigma_{\Delta E}$.  (c) The distribution of total energy change $\Delta E$ (markers).  Solid black line shows the estimate from equation~\ref{eq:heat-dist-cont}.  Inset: on a linear plot, the difference between measurements and equation~\ref{eq:heat-dist-cont} is hardly noticeable.}
\label{fig:VarJar}
\end{figure}

In our experiment, we erase one bit of information, which raises the nonequilibrium free energy of system by $\Delta \mathcal{F}/kT = \ln2$ \cite{parrondo15}.  From the work measurements (figure \ref{fig:pWpQ}~a), we estimate $\sigma_W^2$ and show it versus mean work in figure \ref{fig:VarJar}.   

\subsubsection{Variance of the potential energy difference}

The variance of the potential energy difference is the same for all cycle times $\sigma_{\Delta E}^2 \neq f(\tau)$, since all measurements reported here are done using the same parametrization of the double-well potential and since $\Delta E$ is calculated as the difference between two states that are locally in equilibrium.  We find the difference \mbox{$\Delta E = U(\bar{x}_{N_s+N_0},t_{N_s+N_0})-U(\bar{x}_0,0)$}, where extra time steps ($N_0 = 100 \implies 0.5$~s) are added to ensure that particles have relaxed in the static double-well potential.  

Because all of our protocols start from equilibrium and have an additional relaxation time at the end, the distribution of $\Delta E$ will be the same for all protocol cycle times $\tau$.  We are then able to pool all of our trials together.  We thus histogram all $\Delta E$ to find $p(\Delta E)$, showing the results in figure~\ref{fig:VarJar}c.

To evaluate the distribution $p(\Delta E)$ for a particle in a static double-well potential, we first approximate one well with a harmonic potential of the same curvature around the minima \cite{gavrilov16b}.  The curvature around the minima of potential is given in equation \ref{eq:curvature}.  We approximate each well of a double-well potential with the corresponding local quadratic potential, \mbox{$ U(x)/kT = \frac{1}{2} \kappa x^2 = \frac{1}{2} |\kappa_m| x^2 
	= 4\frac{E_b}{(x_0)^2} (x\pm x_0)^2-12$}. 

The Boltzmann distribution in a harmonic potential implies that in equilibrium, the particle position is distributed as $x \sim \mathcal{N}(0,\tfrac{kT}{\kappa})$, where $\mathcal{N}(0,\sigma^2)$ denotes a Gaussian random variable with mean 0 and variance $\sigma^2$.  For now, we set $\sigma^2=1$ to simplify the analysis.  From the change of variables formula for probability distributions \cite{cowan98}, the random variable $y = x^2$ is distributed as $\chi^2$ distribution with one degree of freedom, $p(y) = \tfrac{1}{\sqrt{2\pi y}} e^{-y/2} \, \theta(y)$, where the Heaviside step function $\theta(y)$ enforces $y \ge 0$.  The characteristic function, $\varphi_0(s) = \langle e^{isy} \rangle$ is then $\varphi_0(s) = \tfrac{1}{\sqrt{1-2is}}$.  The $\Delta E$ distribution is the difference between two such random variables.  If $\tau$ is much longer than the characteristic relaxation time of the trap, the positions at $t=0$ and $t=\tau$ are independent.  In these experiments, the particle relaxes in a static potential for 0.5~s, which is larger than the 0.03~s-long relaxation time of a potential, implying independent positions.  The characteristic function of the difference $y=y_T-y_0$ is then 
\begin{align}
	\varphi(s) = \langle e^{isy} \rangle &= \langle e^{isy_T} \rangle \,  \langle e^{-isy_0} \rangle 
	=\varphi_0(s) \, \varphi_0(-s) = \frac{1}{\sqrt{1+4s^2}} \,,
\end{align}
which is the characteristic function of $p(y) = \tfrac{1}{2\pi} K_0\bigl(\tfrac{|y|}{2} \bigr)$, where  $K_0(x)$ is the modified Bessel function of zero order.  Putting back the variance $\sigma^2$ gives  $p(y) = \tfrac{1}{2\pi \sigma^2} K_0 \bigl(\tfrac{|y|}{2\sigma^2} \bigr)$.  For $\sigma^2 = \bigl( \tfrac{1}{2}\bigr)\kappa \, \bigl( \tfrac{kT}{\kappa} \bigr) = \tfrac{1}{2}kT$, we thus find \cite{imparato07, chatterjee10} 
\begin{align}
	p(\Delta E) = \frac{1}{\pi kT} \, K_0\left(\frac{|\Delta E|}{kT} \right) \,.
\label{eq:heat-dist-cont}
\end{align}

Equation \ref{eq:heat-dist-cont} is shown in figure \ref{fig:VarJar}(c) and compared with measured $p(\Delta E)$.  The general shape is as expected. The discrepancies may be traced back to the finite update time $\Delta t$ of the feedback trap, rather than the deviation of local well shape from a harmonic potential.

\subsubsection{Why the heat distribution is broader than the work distribution}

We are now in a position to understand the perhaps-surprising result that heat measurements in a long protocol have a distribution that is much broader than the corresponding work measurement.  As we have seen (equation~\ref{eq:variances}), the variances are related as $\sigma_W^2 + \sigma_{\Delta E}^2 = \sigma_Q^2$.  But, as figure~\ref{fig:VarJar} a and the constraint imposed by the Jarzynski relation (equation~\ref{eq:GaussCumulantIntro2}) shows, $\sigma_W^2$ \textit{must} go to zero in the long-cycle ($\tau \to \infty$) limit.  Thus, at long times, we expect $\sigma_Q \approx \sigma_E$.  Yet the arguments in the previous suggestion (and equation~\ref{eq:heat-dist-cont}) both show that $\sigma_{\Delta E} \approx kT$, independent of cycle time.  (Recall that the energy is a state variable.)  More physically, even in a static potential, where the work is identically zero, heat always sloshes into and out from the heat bath.  Even for a static potential, the width (equation~\ref{eq:heat-dist-cont}) is of order $kT$.  Thus, the width of the heat distribution tends to $kT$ and not zero for long cycle times.

In many previous experimental studies \cite{toyabe10, roldan14, martinez15, blickle12, koski13, koski14, koski14b, koski15, Hong16, Martini16, Peterson16, rossnage16, berut12, Jun14, berut13, gavrilov16a, proesmans16}, measurements start and end in with the system in local equilibrium, where $p(\Delta E)$, $\langle \Delta E \rangle$, and $\sigma_{\Delta E}$ can be all computed directly from the shape of the potential $U(x)$.  In such cases, which often correspond to slow protocols, the analysis given here suggests that the mean heat estimate $\langle Q \rangle$  should be inferred by combining the measured mean work $\langle W \rangle$ with the computed value of  $\langle \Delta E \rangle$.

\section{Conclusion}

In this paper, we have reviewed the use of feedback traps for studying fundamental problems of nonequilibrium thermodynamics and information theory.  In particular, we repeated and improved  recent experiments on memory erasure that probe the Landauer limit.  The more-efficient protocols developed, while still not optimal \cite{aurell12,zulkowski14}, allowed more detailed exploration than previous work.

We used the extended measurements to compare work and heat measurements.  Although there remain basic mathematical questions as to how best to estimate heat from a stochastic time series, we found---following a suggestion by D. Chiuchi{\`u}---that a higher-order numerical method based on Simpson's rule cures the most basic issue, a secular divergence of the estimate that, because it grows with time, always dominates when applied to long trajectories.  It would be interesting to understand this issue better and to have a more proper integration scheme.

In comparing the work and heat measurements, we found their average values to be approximately consistent with the expected values from the first law of thermodynamics but that the width of the work distribution becomes much narrower than that of the heat distribution, which is always of order  $kT$.  Such behavior is expected for long time series.  It implies that, even if you are interested in estimating the mean heat transferred to a bath (as in these erasure experiments), you will minimize statistical errors by first estimating the mean of the work distribution and then using the first law to convert that estimate into one for the heat transfer.

Finally, although there has not been space to review the work, the feedback trap has also been used for several related projects.  In this paper, we have discussed information erasure assuming that bits are symmetric and associated with equal volumes in phase space, but we can use the unique capabilities of the feedback trap to design potentials where different bits have different sizes \cite{gavrilov16b}.  For example, parts of the potential can be stretched by a controllable factor $\eta$.  Such a situation has practical interest---in biological systems, information-bearing degrees of freedom correspond to different states having different free energies---and also fundamental interest---the asymmetric states are typically not in global equilibrium.  Their exploration further illustrates the dual role of control in cybernetical physics, where it can lead both to new tools and to new physical results.





\begin{acknowledgments}
This work was supported by NSERC Canada.  We thank Davide Chiuchi{\`u} for his suggestions concerning numerical methods and David Sivak for helpful conversations.
\end{acknowledgments}
\bibliographystyle{abbrv}

\end{document}